\documentclass[aps,prx,reprint,superscriptaddress,longbibliography]{revtex4-2}

\usepackage{amsmath,amssymb,bm,mathtools}
\usepackage{siunitx}
\usepackage{physics}
\usepackage{graphicx}
\usepackage{xcolor}
\usepackage[colorlinks=true,linkcolor=blue,citecolor=blue,urlcolor=blue]{hyperref}
\usepackage{multirow}

\newcommand{\IFt}{I_{\mathrm{FT}}}

\newcommand{\Iph}{I_{\mathrm{ph}}}
\newcommand{\Iel}{I_{\mathrm{el}}}
\newcommand{\Imod}{I_{\mathrm{model}}}
\newcommand{\ER}{E_R}
\newcommand{\Lz}{L_{\Gamma}}

\newcommand{\BaselineModel}{\emph{baseline recoil model}}
\newcommand{\ExtendedModel}{\emph{explicit electronic convolution model}}

\begin{document}

\title{Graphene lattice recoil in hard X-ray photoemission: Experiment and Theory}

\author{Simone Ritarossi}
\email{simone.ritarossi@uniroma3.it}
\affiliation{Dipartimento di Scienze, Università degli Studi di Roma Tre, Rome, Italy}
\affiliation{INFN Sezione di Roma Tre, Rome, Italy}

\author{Alice Apponi}
\affiliation{INFN Sezione di Roma Tre, Rome, Italy}

\author{Orlando Castellano}
\affiliation{Dipartimento di Scienze, Università degli Studi di Roma Tre, Rome, Italy}
\affiliation{INFN Sezione di Roma Tre, Rome, Italy}

\author{Jos\'e Lorenzana}
\affiliation{ISC-CNR Institute for Complex Systems and Department of Physics, Sapienza University of Rome, Rome, Italy}

\author{Domenica Convertino}
\affiliation{Center for Nanotechnology Innovation @NEST, Istituto Italiano di Tecnologia, Pisa, Italy}

\author{Camilla Coletti}
\affiliation{Center for Nanotechnology Innovation @NEST, Istituto Italiano di Tecnologia, Pisa, Italy}
\affiliation{Graphene Labs, Istituto Italiano di Tecnologia, Via Morego 30, I-16163 Genova, Italy}

\author{Tien-Lin Lee}
\affiliation{Diamond Light Source Ltd, Harwell Science and Innovation Campus, Didcot, Oxfordshire, United Kingdom}

\author{Francesco Offi}
\affiliation{Dipartimento di Scienze, Università degli Studi di Roma Tre, Rome, Italy}
\affiliation{INFN Sezione di Roma Tre, Rome, Italy}

\author{Alessandro Ruocco}
\affiliation{Dipartimento di Scienze, Università degli Studi di Roma Tre, Rome, Italy}
\affiliation{INFN Sezione di Roma Tre, Rome, Italy}

\date{\today}

\begin{abstract}
Hard-x-ray C~$1s$ photoemission from monolayer graphene probes a regime in which nuclear recoil and intrinsic electronic asymmetry contribute on comparable energy scales to the observed spectral line shape.
Here we combine experiment and modeling over the photon-energy range \SI{0.8}{keV}--\SI{8}{keV} to resolve this interplay quantitatively.
A graphene-specific implementation of the Fujikawa--Takata cumulant formalism, based on an anisotropic vibrational density of states constrained by first-principles phonon calculations, captures the expected recoil scaling with photon energy and emission geometry but fails to reproduce the pronounced asymmetric tails of the measured spectra.
To overcome this limitation, we introduce an explicit electronic convolution model in which an intrinsic, photon-energy-independent electronic line shape extracted from near-recoilless \SI{0.8}{keV} data is convolved with a phonon recoil kernel carrying the full dependence on photon energy and emission angle.
This approach reproduces both the measured line-shape evolution and the observed centroid shifts across the explored energy range without refitting the spectra at higher photon energies.
The results show that recoil in graphene cannot be described by a baseline
treatment in which the phonon recoil kernel is combined only with symmetric
lifetime broadening, but must be treated together with the intrinsic many-body
electronic response of the C~$1s$ line.
\end{abstract}

\maketitle

\section{Introduction}
\label{sec:intro}

At hard-x-ray photon energies \cite{Fadley2010,Fadley2016}, core-level
photoemission enters a regime in which the momentum carried by the
photoelectron is large enough for nuclear recoil to become an intrinsic
part of the excitation process, especially for light elements. In this
regime, the emitting atom cannot be regarded as an infinitely massive
scatterer: the sudden creation and emission process transfers momentum
to the lattice, producing measurable modifications of the spectral
profile, including energy shifts, recoil broadening, and geometry-dependent
line-shape changes.

A quantitative framework for this problem was developed by Fujikawa and
Takata, who formulated the recoil contribution in terms of a cumulant
description of the lattice response \cite{Fujikawa2006,Takata2007}.
In their application to graphite, Takata \textit{et al.} showed that the
C~$1s$ hard-x-ray photoemission line is reshaped by the coupling between
the photoelectron momentum and the anisotropic vibrational motion of
carbon atoms, thereby establishing graphite as a prototypical solid-state
system for recoil spectroscopy \cite{Takata2007}. Direct experimental
evidence of recoil in condensed-matter photoemission was also obtained from the Fermi edge of simple metals, where the displacement and
broadening of the Al edge relative to Au demonstrated that nuclear recoil
is directly observable at hard-x-ray energies \cite{Takata2008}.
Consistent recoil signatures in highly oriented pyrolytic graphite were
also connected to electron scattering and neutron Compton scattering
measurements within the same general physical picture \cite{Vos2008}. Related effects have been identified in molecular core-level spectra,
where the discrete vibrational structure of light molecules makes the
transferred momentum traceable mode by mode \cite{Kukk2018}. Taken
together, these studies show that recoil is not merely an instrumental
or kinematic correction to hard-x-ray photoemission spectra. Rather, for
light-element systems, it is part of the many-body line-shape problem and
must be treated together with intrinsic electronic structure, vibrational
dynamics, and experimental resolution.

\noindent Graphene provides a particularly clean setting for this problem.
Its lattice dynamics is strongly anisotropic, with distinct in-plane and flexural phonon branches \cite{Dubay2003,FALKOVSKY,Viola,Zhao,Taleb,Politano2012,Michel}.
At the same time, the C~$1s$ line shape in $sp^2$ carbon systems is intrinsically asymmetric: the sudden creation of a core hole couples to low-energy electron--hole excitations, producing the characteristic high-binding-energy tail described by the Doniach--\v{S}unji\'c (DS) form \cite{Doniach,NozieresDeDominicis1969}.
This asymmetry was established in graphite by Sette \textit{et al.}~\cite{Sette1990}, who showed that lifetime and screening effects are inseparable in the measured line shape and must be modeled explicitly.
More recently, soft-x-ray photoemission on graphene membranes confirmed that the asymmetric tail survives in suspended graphene and is therefore an intrinsic feature of the graphene C~$1s$ spectrum \cite{Apponi2024}.

\noindent In the hard-x-ray regime, recoil and intrinsic asymmetry act on comparable energy scales.
As the photoelectron kinetic energy increases, the free-atom recoil energy
\begin{equation}
\ER=\frac{\hbar^2 k^2}{2M_C}=\frac{m_e}{M_C}\,E_{\mathrm{kin}},
\label{eq:ER_intro}
\end{equation}
grows linearly with $E_{\mathrm{kin}}$.
For carbon, $\ER$ reaches approximately \SI{172}{meV} at \SI{4}{keV} and \SI{356}{meV} at \SI{8}{keV}, values large enough to compete with the intrinsic linewidth of the C~$1s$ line and to reshape the observed spectrum in a geometry-dependent manner.
The resulting problem is therefore not simply one of recoil broadening, but of disentangling phonon-driven spectral redistribution from an already asymmetric intrinsic electronic response.

\noindent In this work we address this problem through a combined experimental and theoretical study of angle- and photon-energy-dependent C~$1s$ photoemission from clean monolayer graphene extending into the hard-x-ray regime.
We first adapt the Fujikawa--Takata cumulant formalism \cite{Fujikawa2006,Takata2007,Takata2008,Kukk2018,Kayanuma2016}, originally developed for graphite, to a graphene-specific anisotropic vibrational density of states.
We then show that a baseline recoil treatment in which nonphononic effects are
represented only by a symmetric Lorentzian lifetime term is insufficient to
account for the measured spectra, and introduce an explicit convolution scheme
in which the intrinsic electronic line shape is determined independently from
near-recoilless data.
Section~\ref{sec:Experiment} presents the experiment.
Sections~\ref{sec:FT_framework} summarize the cumulant framework and its graphene-specific anisotropic vibrational DOS.
Section~\ref{sec:total_model} introduces the explicit electronic convolution model and compares it with experiment.

\section{Experiment}
\label{sec:Experiment}

A photoemission spectroscopy investigation of the C~$1s$ core-level was conducted on monolayer graphene in the soft-to-hard X-ray regimes. The experiment was performed at the I09 beamline of Diamond Light Source \cite{Lee2018}, with photon energies of 801.02, 4053.4, 8079.2~eV (respectively called 0.8, 4, 8~keV from here on). The I09 experimental chamber is equipped with a VG Scienta EW400 HAXPES analyzer and the spectral resolution at each photon energy is 0.14~eV (0.8~keV), 0.24~eV (4~keV) and 0.30~eV (8~keV) -- as measured from the Fermi edge of a polycrystalline Au reference sample mounted in electrical and thermal contact with the graphene one. The Fermi edge of the Au foil was also exploited to calibrate the binding energy scale.

\noindent The investigation was conducted on a sample of polycrystalline monolayer graphene grown on electropolished copper via chemical vapor deposition \cite{Miseikis_2015, Convertino2020} then transferred onto a commercial nickel grid (G2000HAN - Ted Pella Inc.) with the standard wet etching technique \cite{Miseikis_2015, Li2009, D1CP04316A} -- further details on the sample preparation can be found in \cite{Apponi2024}. Note that the grids employed for transfer are metallic only with no additional lacey carbon film or mesh, in order to avoid superimposition of spurious C~$1s$ signals. Following preparation, the graphene sample underwent high-temperature annealing (650~$^\circ$C) in ultra-high vacuum ($\sim10^{-10}$~mbar) at the LASEC laboratory of Universit\`{a} Roma Tre, in order to remove contaminants and polymeric residuals of the transfer procedure \cite{Apponi2024}. The sample was then transported in low-vacuum ($\sim 10^2$~mbar) to the I09 beamline, where a further annealing in ultra-high vacuum was performed at 250~$^\circ$C to remove adventitious contaminants eventually adsorbed during the in-air mounting of the sample in the I09 apparatus.

\noindent The C~$1s$ core-level was measured at room temperature ($T=\SI{300}{K}$) from the clean monolayer graphene on grid sample with increasing photon energies while keeping the experimental geometry fixed. In particular, the experiment was conducted at grazing incidence -- 24$^\circ$ with respect to the surface -- enhancing the sensitivity to the single graphene layer. By exploiting the angular acquisition characteristic of the electron analyzer, the spectra were recorded at both 90$^\circ$ and 45$^\circ$ emission angles to reveal differences in the phononic excitations driven by recoil mechanisms. At the investigated energies, the photon spot size on the surface is on the order of a few hundred $\mu m^2$. Given the 8~$\mu m$ diameter of the grid holes, the measured C~$1s$ signal includes contribution from graphene both suspended over the holes and supported by the nickel grid.
\begin{figure}[h!]
  \centering
  \includegraphics[width=\linewidth]{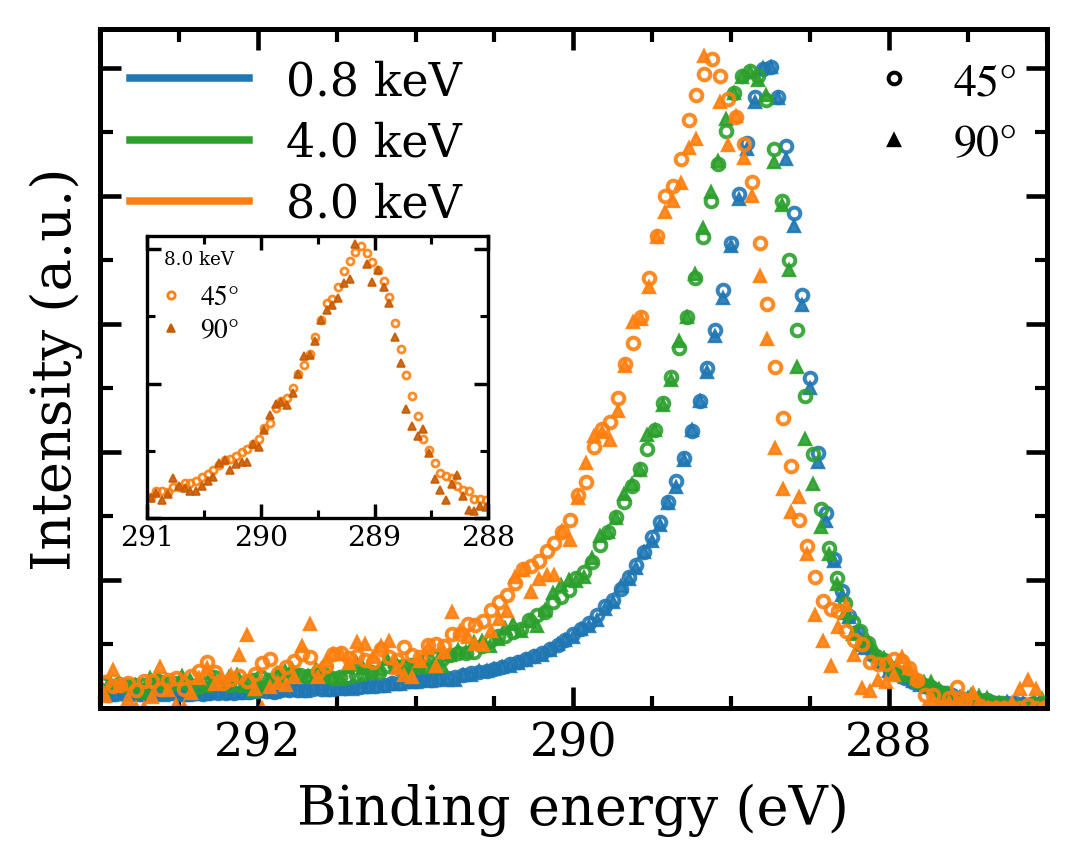}
  \caption{Experimental graphene C~$1s$ spectra at $T=\SI{300}{K}$ for increasing photon energy, as labeled, and for two emission angles, $45^\circ$ (open circles) and $90^\circ$ (triangles), measured from the surface normal. The inset shows an enlarged view of the \SI{8}{keV} spectra, where the recoil-induced anisotropy between the two emission geometries is most clearly visible.}
  \label{fig:data}
\end{figure}
\noindent Prior to the spectral analysis, an iterative Shirley-type background \cite{Shirley1972} was subtracted from all measured spectra to remove the inelastic scattering contribution. The measured C~$1s$ spectra at increasing photon energy (0.8, 4 and 8~keV) are reported in Figure \ref{fig:data}, for emission angles of 45$^\circ$ and 90$^\circ$. The spectra exhibit an asymmetric line shape at all photon energies, indicating the intrinsic electronic contribution associated with the metallic character of graphene. Furthermore, with increasing photon energy the recoil effects induce two distinct signatures: a shift of the peak and a broadening of the spectrum toward higher binding energies. The recoil-driven modifications of the line shape show also an anisotropic behavior at 45$^\circ$ and 90$^\circ$, enhanced at higher photon energy. The anisotropy reflects the different relative weight of in-plane and flexural phononic excitations depending on the electron emission angle, as will be thoroughly discussed in Section~\ref{sec:FT_framework}.

\section{Fujikawa--Takata framework applied to monolayer graphene}
\label{sec:FT_framework}

The lattice contribution to recoil in core-level photoemission is described using the cumulant formalism introduced by Fujikawa \cite{Fujikawa2006} and extended to anisotropic phonon baths by Takata \cite{Takata2007}; related cumulant approaches for phonon contributions to electron spectral functions have also been developed in other contexts \cite{Story2014}, and a broader overview is given in Ref.~\cite{Kayanuma2016}.
In the following we summarize the expressions needed for the present analysis and emphasize how the momentum transfer couples to lattice displacements and determines the recoil response of the measured spectrum.

\noindent The formal starting point is the Born--Oppenheimer approximation, which separates electronic and nuclear degrees of freedom on the basis of their widely different timescales.
Within this approximation, the core orbital remains centered on the instantaneous nuclear position of the emitting atom, and the photoemission matrix element factorizes into an electronic contribution, which encodes the intrinsic line shape, and a lattice contribution, which depends only on the nuclear coordinates.
This separation underlies the convolution structure of the full spectrum developed below.

\noindent The nuclear coordinate operator is written as $\mathbf{R}=\mathbf{R}_0+\mathbf{u}$, where $\mathbf{R}_0$ is the equilibrium lattice site and $\mathbf{u}$ is the fluctuating part of the atomic position.
Thermal and zero-point lattice vibrations correspond to fluctuations of $\mathbf{u}$, such that $\langle \mathbf{u}\rangle_{\beta}=0$ while $\langle \mathbf{u}^2\rangle_{\beta}\neq 0$, where $\langle \cdots \rangle_{\beta}$ denotes a quantum and canonical thermal average at inverse temperature $\beta=(k_{\mathrm B}T)^{-1}$.
The time dependence relevant to recoil is governed by the vibrational Hamiltonian $H_{\mathrm{vib}}$ and is written in the Heisenberg picture as
\begin{equation}
\mathbf{u}(t)=e^{iH_{\mathrm{vib}}t/\hbar}\,\mathbf{u}\,e^{-iH_{\mathrm{vib}}t/\hbar}.
\end{equation}

\noindent Recoil is controlled by final the momentum $\hbar\mathbf{k}$ of the photoelectron.
To leading order, $\mathbf{k}$ is fixed by the photoelectron kinetic energy and emission geometry.
Corrections arising from multiple scattering (photoelectron diffraction) have been shown to amount to only a few meV for graphite-like carbon at multi-keV kinetic energies \cite{Fujikawa2008} and are therefore neglected here.
Within the adiabatic approximation, the C \(1s\) core orbital is assumed to follow the instantaneous position of the emitting carbon atom. Thus, if \(\mathbf{R}=\mathbf{R}_0+\mathbf{u}\), the core-electron wave function is taken as $\langle \mathbf{r}|\psi_c\rangle = \psi_c(\mathbf{r}-\mathbf{R})$. After projection onto the outgoing photoelectron state, this rigid translation of the core orbital produces the recoil phase factor associated with the nuclear displacement. Neglecting the photon momentum with respect to the photoelectron momentum, the phonon part of the spectrum is then governed by the correlation function
\[
F(t)
=
\left\langle
e^{i\mathbf{k}\cdot\mathbf{u}(t)}
e^{-i\mathbf{k}\cdot\mathbf{u}(0)}
\right\rangle ,
\]
which encodes the lattice response to the impulsive momentum transfer,
defined as a canonical average at inverse temperature $\beta=(k_B T)^{-1}$.
Its structure is closely related to the time-dependent pair correlation function introduced by van~Hove \cite{vanHove1954} in the context of neutron scattering.
The corresponding phonon recoil kernel in the energy domain is obtained from the Fourier transform of $F(t)$; up to an overall prefactor,
\begin{equation}
\Iph(E)\propto \int_{-\infty}^{+\infty}dt\;
\exp\!\left(-\frac{i}{\hbar}Et\right)\,F(t).
\label{eq:Iph}
\end{equation}
Here $E$ is the binding-energy coordinate measured relative to a fixed recoilless reference,
\begin{equation}
E \equiv E_B - E_B^{(0)},
\label{eq:E_def}
\end{equation}
where $E_B^{(0)}$ is determined from the near-recoilless \SI{0.8}{keV} spectrum.

\noindent In the original Fujikawa--Takata implementation \cite{Takata2007}, homogeneous broadening associated with the core-hole lifetime is modeled by an additional exponential factor in the time domain, with $\Gamma=0.16\,\mathrm{eV}$:
\begin{equation}
\IFt(E)\propto \int_{-\infty}^{+\infty}dt\;
\exp\!\left(-\frac{i}{\hbar}Et\right)\,
\exp\!\left(-\frac{\Gamma}{\hbar}\abs{t}\right)\,
F(t).
\label{eq:I_FT}
\end{equation}
Equivalently, multiplication by $\exp[-(\Gamma/\hbar)\abs{t}]$ corresponds in the energy domain to a convolution with a Lorentzian kernel $\Lz$,
\begin{equation}
\IFt(E)\propto \qty[\Lz \ast \Iph](E),
\label{eq:Lorentz_conv}
\end{equation}
where $\ast$ denotes convolution.
In what follows, the formulation of Eqs.~\eqref{eq:I_FT}--\eqref{eq:Lorentz_conv}
is referred to as the \BaselineModel: a recoil cumulant treatment in which
nonphononic effects are represented by a symmetric homogeneous lifetime
broadening.
This terminology does not imply that the phonon recoil kernel itself is
strictly symmetric; rather, it emphasizes that the baseline model contains no
explicit asymmetric electronic line shape.
As shown for graphite in Ref.~\cite{Takata2007}, this baseline approach
reproduces the characteristic recoil scaling of spectral displacement and
overall width with photoelectron energy and emission geometry. \\

The Fujikawa--Takata formalism is adapted here to monolayer graphene by introducing a graphene-specific anisotropic vibrational DOS.
In the harmonic approximation, lattice displacements are Gaussian and the linked-cluster expansion of the generating function terminates exactly at second order.
Accordingly, $F(t)=\exp[G(t)]$, with
\begin{equation}
G(t)=-\frac{1}{2}\left\langle \bigl(\mathbf{k}\cdot[\mathbf{u}(t)-\mathbf{u}(0)]\bigr)^2\right\rangle_{\!\beta}.
\label{eq:cumulant_gaussian}
\end{equation}

\noindent Graphene exhibits strongly anisotropic lattice dynamics, with qualitatively different in-plane and flexural excitations.
In the cumulant formulation, the emission-angle dependence becomes explicit in a mode-resolved representation: decomposing the lattice response into in-plane and out-of-plane channels yields the geometric weights $\cos^2\theta$ and $\sin^2\theta$, where $\theta$ denotes the photoelectron emission angle with respect to the sample surface.
The mode decomposition is written as
\begin{equation}
G(t)=
\int_{-\infty}^{+\infty} d\omega\;\bigl(e^{i\omega t}-1\bigr)\,
\Bigl[
J_s(\omega)\cos^2\theta
+
J_b(\omega)\sin^2\theta
\Bigr].
\label{eq:G_aniso}
\end{equation}
Stretching ($\lambda=s$) and bending ($\lambda=b$) excitations are treated as two bosonic channels, respectively in-plane and out-of-plane channels, and the cumulant is expressed in terms of channel-resolved spectral densities $J_\lambda(\omega)$ constructed from a normalized vibrational DOS $D_\lambda(\omega)$.
Introducing the Bose occupation $n(\omega)=\bigl(e^{\beta\hbar\omega}-1\bigr)^{-1}$ for $\omega>0$ and the step function $\Theta(\omega)$, one obtains, in direct analogy with Ref.~\cite{Takata2007},
\begin{equation}
J_\lambda(\omega)
=
\frac{\ER}{\hbar\,\omega}\,
D_\lambda(\abs{\omega})\,
\Bigl\{
\bigl[n(\abs{\omega})+1\bigr]\Theta(\omega)
+
n(\abs{\omega})\Theta(-\omega)
\Bigr\}.
\label{eq:J_from_D}
\end{equation}
The two terms in braces correspond to phonon emission ($\omega>0$) and phonon
absorption ($\omega<0$), respectively.
Because these contributions are weighted by $n(\abs{\omega})+1$ and
$n(\abs{\omega})$, the phonon recoil kernel is not constrained to be strictly
symmetric.
For the present room-temperature spectra, however, the asymmetry generated by
the phonon kernel, after lifetime and instrumental broadening, is not sufficient
to account for the pronounced high-binding-energy tail observed experimentally.

\noindent The free-atom recoil scale in Eq.~\eqref{eq:ER_intro} provides an immediate kinematic estimate of the recoil energy and implies a linear increase of recoil distortions with $E_{\mathrm{kin}}$.
Within the cumulant construction, this is reflected by the proportionality $J_\lambda(\omega)\propto \ER$, so that the strength of the phonon recoil response grows systematically with photoelectron kinetic energy.
All quantities defined so far depend parametrically on $(E_{\mathrm{kin}},\theta,T)$; for brevity these arguments are left implicit below, although this dependence is central to the predictive content of the model.

\begin{figure}[h!]
  \centering
  \includegraphics[width=\linewidth]{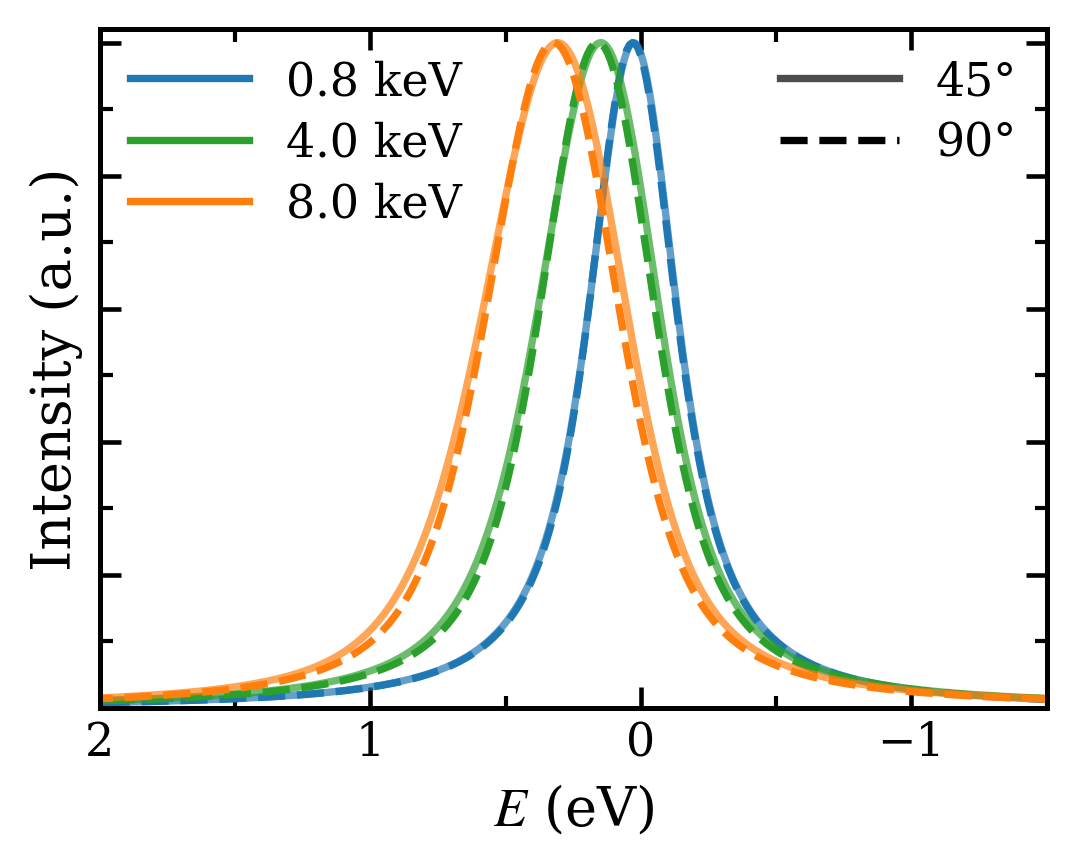}
  \caption{Theoretical graphene C~$1s$ spectra at $T=\SI{300}{K}$ computed within the \BaselineModel\ [Eq.~\eqref{eq:I_FT}] for the photon energies indicated in the figure and for emission angles of $45^\circ$ (solid lines) and $90^\circ$ (dashed lines).}
  \label{fig:basic_model}
\end{figure}

\noindent To construct an anisotropic vibrational DOS for graphene, we use the long-wavelength behavior of the acoustic phonons.
The in-plane acoustic branches are approximately linear, $\omega\simeq v q$, whereas the flexural ZA branch is quadratic, $\omega \propto q^2$, as expected for an elastic two-dimensional sheet and confirmed by density functional perturbation theory (DFPT) phonon dispersions \cite{MarianiOppen2007,Yan2008,Michel,Mounet2005}.
In two dimensions, these dispersions imply $D(\omega)\propto \omega$ for linear branches and an approximately constant DOS for a quadratic branch.
Accordingly, we adopt the analytic form
\begin{equation}
D_\lambda(\omega)=
\begin{cases}
\dfrac{2\omega}{\omega_s^2}, & \lambda=s,\ \ 0<\omega<\omega_s,\\[6pt]
\dfrac{1}{\omega_b}, & \lambda=b,\ \ 0<\omega<\omega_b,\\[6pt]
0, & \text{otherwise},
\end{cases}
\label{eq:Dlambda_forms}
\end{equation}
with cutoff energies fixed from density functional pertubation theory (DFPT) by the upper edge of the relevant phonon branches:
$\hbar\omega_b\simeq \SI{66}{meV}$ and $\hbar\omega_s\simeq \SI{160}{meV}$ \cite{NikaBalandin2012,Yan2008,Koukaras,Michel}.

\noindent Figure~\ref{fig:basic_model} illustrates the \BaselineModel\ applied to monolayer graphene.
As $E_{\mathrm{kin}}$ increases, the recoil scale $\ER\propto E_{\mathrm{kin}}$ grows, producing both a systematic displacement of the spectrum and a progressive increase in its overall width.
A second key feature is the dependence on emission geometry: changing $\theta$ modifies the relative weight of in-plane and out-of-plane excitations through Eq.~\eqref{eq:G_aniso}, leading to distinct spectral evolutions that directly reflect the anisotropic lattice dynamics of graphene.
Overall, the \BaselineModel\ captures the basic recoil trends expected for
monolayer graphene.
However, as seen in Fig.~\ref{fig:data}, the experimental spectra display a
pronounced asymmetric tail on the high-binding-energy side that cannot be
accounted for by a recoil treatment in which the intrinsic electronic
contribution is reduced to a symmetric Lorentzian lifetime broadening.
This asymmetry originates from the many-body electronic response of the
$sp^2$ carbon lattice and therefore requires an explicit electronic line shape
beyond the single-Lorentzian representation used in the baseline model.

\section{Explicit electronic spectrum: convolution strategy}
\label{sec:total_model}

The intrinsic C~$1s$ line shape of graphene is asymmetric: metallic screening by the $\pi$ electrons generates a continuum of low-energy electron--hole excitations, giving rise to the high-binding-energy tail captured analytically by the DS form \cite{Doniach,NozieresDeDominicis1969}.
The interplay between such many-body electronic excitations and nuclear recoil was first discussed by Hedin \cite{Hedin1980} in the context of shake-up spectra in metals.
In graphene at hard-x-ray energies, phonon-driven recoil superimposes additional spectral displacements and broadenings on top of this asymmetric electronic profile.
A realistic description therefore requires the electronic and phononic responses to be treated as distinct contributions.

\noindent The adiabatic (Born--Oppenheimer) approximation introduced in Sec.~\ref{sec:FT_framework} provides the formal basis for this separation.
Because electronic excitations occur on timescales of order \SI{e-16}{s}, much shorter than the \SI{e-13}{s} scale of nuclear motion, the electronic and phononic responses decouple in time and their contributions to the photoemission cross section factorize.
In the energy domain, this temporal factorization becomes a convolution:
\begin{equation}
\Imod(E)
=
\bigl[\Iel(E)\ast \Iph(E)\bigr]\ast R(E),
\label{eq:Imodel_conv}
\end{equation}
where $\Iel(E)$ is the intrinsic electronic line shape, taken to be photon-energy independent and determined from the near-recoilless \SI{0.8}{keV} data; $\Iph(E)$ is the phonon recoil kernel of Eq.~\eqref{eq:Iph}, carrying the full dependence on $(E_{\mathrm{kin}},\theta,T)$; and $R(E)$ is a Gaussian instrumental resolution function with FWHM fixed to \SI{0.14}{eV} at \SI{0.8}{keV}, \SI{0.24}{eV} at \SI{4}{keV}, and \SI{0.30}{eV} at \SI{8}{keV}.
The separation into $\Iel$ and $\Iph$ is therefore not merely phenomenological, but follows directly from the separation of electronic and nuclear timescales within the Born--Oppenheimer framework.

\begin{figure}[h!]
  \centering
  \includegraphics[width=\linewidth]{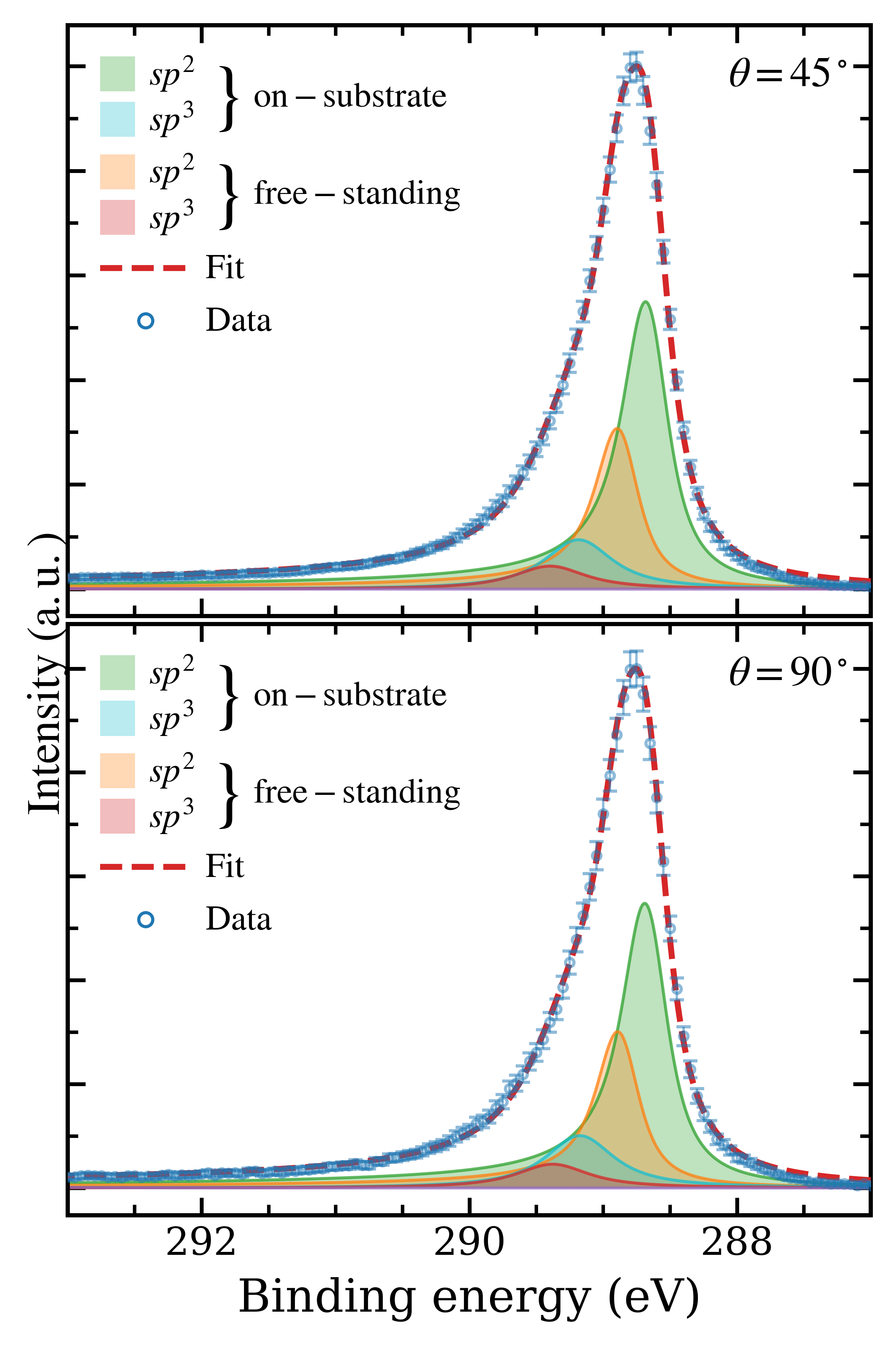}
  \caption{Near-recoilless reference spectrum at \SI{0.8}{keV} and $T=\SI{300}{K}$ shown on the binding-energy scale for emission angles $\theta=45^\circ$ (top) and $\theta=90^\circ$ (bottom). Open circles are the experimental data; the dashed red curve is the best fit obtained by representing the intrinsic electronic line shape as the weighted sum of four Doniach--\v{S}unji\'c components ($sp^2$-like and $sp^3$-like, in both supported and free-standing regions) plus a Gaussian plasmon contribution. Shaded areas indicate the individual components visible within the displayed energy window; the plasmon contribution is included in the fit but lies outside the plotted range. Best-fit parameters are reported in Table~\ref{tab:recoilless_fit_params}.}
\label{fig:0.8}
\end{figure}

\noindent The intrinsic electronic spectrum $\Iel(E)$ entering Eq.~\eqref{eq:Imodel_conv} is determined from the near-recoilless reference shown in Fig.~\ref{fig:0.8}.
We parameterize it as the weighted sum of four DS components plus a broad Gaussian plasmon-loss feature (note that this plamon loss feature is outside the energy range shown in Fig.~\ref{fig:0.8}, being at 6 eV higher binding energy than the main peak).
The DS form captures asymmetric core-level line shapes arising from many-body screening and low-energy electron--hole excitations \cite{Doniach}, while the Gaussian term accounts for additional incoherent loss weight at larger energy, as expected for plasmon-assisted photoemission \cite{Politano2017,Yuan2011,Guzzo2014,Leiro2003}.
Guided by the coexistence, within the probed area, of regions that are effectively free-standing and regions weakly coupled to the substrate, we include two components for each environment.
These are labeled $sp^2$ and $sp^3$ for clarity: $sp^2$ denotes the graphitic-like contribution characteristic of trigonal carbon in graphene, whereas $sp^3$ is used more broadly as a convenient label for locally perturbed carbon environments associated with a more symmetric line shape.
This yields a minimal yet flexible representation of the reference spectrum.
Although four DS functions are included in the fit, two of them converge to $\alpha=0$ and therefore reduce to purely Lorentzian line shapes; these are identified with the $sp^3$-like contributions, while the remaining two retain finite asymmetry and describe the $sp^2$-like components.
The best-fit parameters are summarized in Table~\ref{tab:recoilless_fit_params}.
Importantly, they remain essentially unchanged between $\theta=45^\circ$ and $\theta=90^\circ$, supporting the use of a single angle-independent $\Iel(E)$ determined at \SI{0.8}{keV} and reused unchanged at higher photon energies.

\begin{table}[t]
\caption{\label{tab:recoilless_fit_params}
Best-fit parameters for the near-recoilless reference spectrum at \(h\nu=\SI{0.8}{keV}\).
The $w_j$ are relative weights (normalized within each fit), $s_j$ are component positions, and $\alpha_j$ is the Doniach--\v{S}unji\'c asymmetry parameter ($\alpha_j=0$ corresponds to the Lorentzian limit).
The reduced chi-square $\chi_\nu^2$ is reported for each emission angle.}
\begin{ruledtabular}
\footnotesize
\setlength{\tabcolsep}{3pt}
\renewcommand{\arraystretch}{1.05}
\begin{tabular}{c l c c c c}
$\theta$ & Component & $w_j$ & $s_j$ (eV) & $\Gamma_j$/FWHM (eV) & $\alpha_j$ \\
\hline
\multirow{5}{*}{$45^\circ$}
& $sp^2$  & 0.543 & -0.14 & 0.16 & 0.10 \\
& $sp^3$  & 0.097 & 0.37 & 0.29 & 0 \\
& $sp^2$  & 0.310 & 0.07 & 0.14 & 0.10 \\
& $sp^3$  & 0.046 & 0.57 & 0.31 & 0 \\
& Plasmon & 0.004 & 6.01 & 2.42 & -- \\
\multicolumn{1}{r}{$\chi_\nu^2=1.18$ }\\
\hline
\multirow{5}{*}{$90^\circ$}
& $sp^2$  & 0.549 & -0.14 & 0.16 & 0.09 \\
& $sp^3$  & 0.100 & 0.37 & 0.28 & 0 \\
& $sp^2$  & 0.301 & 0.07 & 0.14 & 0.09 \\
& $sp^3$  & 0.046 & 0.57 & 0.30 & 0 \\
& Plasmon & 0.005 & 6.02 & 2.60 & -- \\
\multicolumn{1}{r}{$\chi_\nu^2=1.12$} \\
\end{tabular}
\end{ruledtabular}
\end{table}

\noindent Once determined from the \SI{0.8}{keV} reference, $\Iel(E)$ is kept strictly fixed at all higher photon energies and emission geometries: no refitting of electronic parameters is performed at \SI{4}{keV} or \SI{8}{keV}.
This strategy is justified by the fact that the recoil scale at \SI{0.8}{keV} is only $\ER\simeq\SI{20}{meV}$, so recoil is negligible on the linewidth scale and the measured profile is dominated by intrinsic electronic and instrumental broadening.
With $\Iel(E)$ held fixed, the full energy- and angle-dependent evolution of $\Imod(E)$ is predicted entirely by the phonon recoil kernel $\Iph(E)$, while the instrumental resolution $R(E)$ is fixed independently by the measured FWHM at each photon energy.

\begin{figure}[h!]
  \centering
  \includegraphics[width=\linewidth]{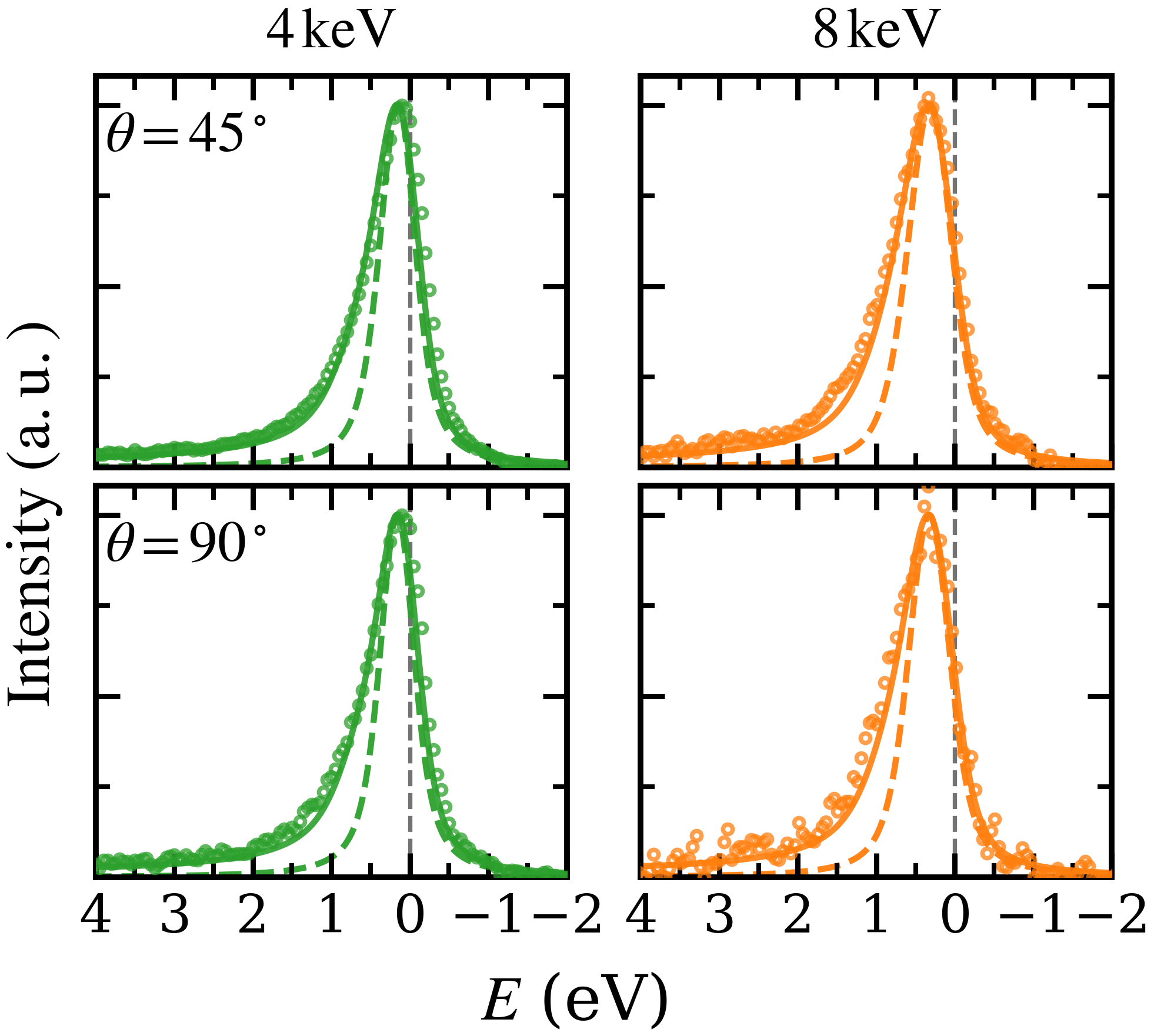}
  \caption{Comparison between experimental C~$1s$ spectra (open circles) and theoretical predictions at $T=\SI{300}{K}$ for emission angles $\theta=45^\circ$ (left column) and $\theta=90^\circ$ (right column), at \SI{4}{keV} (top row) and \SI{8}{keV} (bottom row). The solid red curves show the \ExtendedModel\ [Eq.~\eqref{eq:Imodel_conv}] and the dashed blue curves show the \BaselineModel\ [Eq.~\eqref{eq:I_FT}]. All spectra are displayed on the recoil-energy scale relative to the near-recoilless \SI{0.8}{keV} reference. The \ExtendedModel\ reproduces the overall recoil-induced spectral evolution, including the centroid displacements reported in Table~\ref{tab:centroid_shifts}, the progressive broadening with increasing kinetic energy, and the asymmetric tail on the high-binding-energy side. By contrast, the \BaselineModel\ captures the gross recoil trends but, lacking an explicit asymmetric electronic component, underestimates the high-binding-energy asymmetry at all energies. No free parameters are adjusted at \SI{4}{keV} or \SI{8}{keV}; the only fitted quantities are the DS parameters of Table~\ref{tab:recoilless_fit_params}, determined exclusively from the \SI{0.8}{keV} reference.}
  \label{fig:model_data}
\end{figure}

\noindent The quality of the model predictions across all energies and emission geometries is illustrated in Fig.~\ref{fig:model_data}, while the corresponding centroid displacements are summarized in Table~\ref{tab:centroid_shifts}.
When expressed relative to the near-recoilless \SI{0.8}{keV} reference, the free-atom kinematic estimate yields recoil shifts of approximately \SI{149}{meV} at \SI{4}{keV} and \SI{333}{meV} at \SI{8}{keV}.
Both the \BaselineModel\ and the \ExtendedModel\ reproduce the systematic increase of the spectral displacement with photon energy as well as its angular dependence, with indeed slightly but systematically larger centroid shifts at $\theta=90^\circ$ than at $\theta=45^\circ$.
This trend is consistent with the anisotropic lattice response encoded in Eq.~\eqref{eq:G_aniso}, which changes the relative weight of in-plane and flexural phonon channels with emission geometry.

\noindent Quantitatively, however, the two models differ substantially.
At \SI{4}{keV}, the experimental centroid shifts, about \SI{137}{meV} and \SI{139}{meV} at $45^\circ$ and $90^\circ$, lie slightly below the kinematic estimate.
The \BaselineModel\ gives somewhat smaller values, whereas the \ExtendedModel\ reproduces the measured displacements more closely at both angles.
The difference becomes more pronounced at \SI{8}{keV}, where the experimental centroid shifts reach approximately \SI{321}{meV}--\SI{323}{meV}.
Here the \BaselineModel\ underestimates the observed displacement by several tens of meV, while the \ExtendedModel\ remains within about \SI{10}{meV} of experiment.
These results show that the recoil-induced redistribution of spectral weight cannot be captured quantitatively by a symmetric Lorentzian broadening alone, and that an explicit treatment of the intrinsic electronic line shape is already required in the intermediate-energy regime and becomes essential at the highest photon energies explored here.

\noindent Beyond the centroid displacements, the \ExtendedModel\ also provides a compelling account of the full line shapes shown in Fig.~\ref{fig:model_data}.
Whereas the \BaselineModel\ yields profiles with insufficient
high-binding-energy asymmetry, the convolution strategy naturally transfers the
intrinsic DS asymmetry of $\Iel(E)$ into the full spectrum at every photon
energy.
The model thus captures simultaneously the overall spectral displacement, the progressive broadening with increasing $E_{\mathrm{kin}}$, and the persistent asymmetric tail, all without introducing any additional free parameters beyond those determined from the \SI{0.8}{keV} reference.

\begin{table}[h!]
\caption{\label{tab:centroid_shifts}
Centroid displacements (meV) on the relative energy scale, reported as $E_B-E_B^{(0)}$ for each column.
Here $E_B^{(0)}$ denotes the corresponding \SI{0.8}{keV} reference centroid (kinematic estimate, model, or data).}
\begin{ruledtabular}
\begin{tabular}{c c cc cc cc}
Energy &
\multicolumn{1}{c}{Kinematic} &
\multicolumn{2}{c}{Baseline} &
\multicolumn{2}{c}{Conv.} &
\multicolumn{2}{c}{Data} \\
\cline{3-4}\cline{5-6}\cline{7-8}
& estimation & $45^\circ$ & $90^\circ$ & $45^\circ$ & $90^\circ$ & $45^\circ$ & $90^\circ$ \\
\hline
4 keV   & 149 & 129 & 131 & 141 & 142 & 137 & 139 \\
8 keV   & 333 & 287 & 295 & 313 & 314 & 321 & 323 \\
\end{tabular}
\end{ruledtabular}
\end{table}

\section{Conclusions}
\label{sec:conclusions}

We have investigated hard-x-ray C~$1s$ photoemission from clean monolayer graphene over the photon-energy range \SI{0.8}{keV}--\SI{8}{keV}, combining experiment with a recoil framework that explicitly retains the intrinsic electronic asymmetry of the graphene line shape.

\noindent From a theoretical point of view, the central result is that the spectral
evolution of graphene in the multi-keV regime cannot be accounted for by a
baseline recoil treatment in which nonphononic effects are represented only by
symmetric lifetime broadening.
A quantitative description requires two distinct ingredients: an anisotropic phonon recoil kernel, which controls the energy- and angle-dependent redistribution of spectral weight, and an intrinsic electronic line shape, which fixes the asymmetric character of the spectrum.
Within the adiabatic framework, these two contributions remain separable and combine naturally through convolution in the energy domain.

\noindent Based on this separation, we introduced an \ExtendedModel\ in which the intrinsic electronic spectrum $\Iel(E)$ is determined once from the near-recoilless \SI{0.8}{keV} data and then kept fixed at higher photon energies, while the phonon recoil kernel $\Iph(E)$ carries the full dependence on photon energy, emission geometry, and temperature through a graphene-specific anisotropic vibrational DOS.
Without refitting at \SI{4}{keV} and \SI{8}{keV}, this approach reproduces both the measured centroid displacements and the full line-shape evolution across the explored energy and angular range, whereas the \BaselineModel\ systematically underestimates the spectral displacement at the highest photon energy and fails to reproduce the pronounced asymmetric tail.

\noindent More broadly, these results show that recoil in graphene is not a simple rigid shift of the C~$1s$ line, but a redistribution of spectral weight governed by the interplay between lattice dynamics and many-body electronic response.
The convolution strategy developed here therefore provides a physically
grounded route to the analysis of recoil effects in other two-dimensional
$sp^2$ systems and, more generally, in materials where electronic and phononic
timescales remain well separated and a near-recoilless reference spectrum can be
identified.
The same framework may also be extended to more complex lattices with inequivalent sublattices, where mass-dependent recoil decompositions of the spectral line shape have been predicted \cite{Kayanuma2011}.

\begin{acknowledgments}
We acknowledge financial support under the National Recovery and Resilience Plan (NRRP), Mission 4, Component 2, Investment 1.1, Call for tender No. 104 published on 2 February 2022 by the Italian Ministry of University and Research (MUR), funded by the European Union -- NextGenerationEU -- Project Title PACE (20227F53E4), CUP F53C24000790006, Grant Assignment Decree No. 20429 adopted on 6 November 2024 by the Italian Ministry of University and Research (MUR). This work was partially supported by the Italian Ministry of University and Research (MUR) under the Grant of Excellence Departments, Art. 1, commi 314--337, Law 232/2016, to the Department of Science, Roma Tre University. The support from Diamond Light Source,
instrument I09 (proposal 38172), is gratefully acknowledged.
\end{acknowledgments}

\section*{Author Contributions}
S.R.: Methodology, Data curation, Software, Formal analysis, Visualization, Writing – original draft, Writing – review and editing. A.A.: Conceptualization, Investigation, Data curation, Formal analysis, Writing – original draft, Writing – review and editing. O.C.: Investigation, Data curation, Formal analysis, Writing – review and editing. J.L.: Supervision, Writing –review and editing. D.C: Resources.Writing –review and editing. C.C.: Resources, Funding acquisition, Writing –review and editing. F.O.:  Investigation, Data Curation, Supervision,  Writing – review and editing. T.-L.L.:Resources.Writing –review and editing. A.R.: Conceptualization, Investigation, Supervision, Funding acquisition, Writing – review and editing.

\section*{Data Availability}

The data used in this work are
available from the corresponding author upon reasonable request.

\bibliographystyle{apsrev4-2}
\bibliography{cit}

\end{document}